\documentclass[psfig,epsfig,final]{article}
\usepackage{sw20aip}

\newcommand{\be}{\begin{equation}}
\newcommand{\ee}{\end{equation}}
\newcommand{\bea}{\begin{eqnarray}}
\newcommand{\eea}{\end{eqnarray}}

\input{tcilatex}
\begin{document}

\title{Predicting Ecosystem Response to Perturbation from Thermodynamic Criteria}
\author{K. Michaelian, V. Alonso Ch\'avez \\
Depto. F\'{i}sica Experimental, Instit\'uto de F\'{i}sica, UNAM, \\
\indent  C.P. 04510, M\'exico D.F., M\'exico}
\maketitle

\begin{abstract}
The response of ecosystems to perturbations is considered from a
thermodynamic perspective by acknowledging that, as for all macroscopic
systems and processes, the dynamics and stability of ecosystems is subject
to definite thermodynamic law. For open ecosystems, exchanging energy, work,
and mass with the environment, the thermodynamic criteria come from
non-equilibrium or irreversible thermodynamics. For ecosystems during
periods in which the boundary conditions may be considered as being
constant, it is shown that criteria from irreversible thermodynamic theory
are sufficient to permit a quantitative prediction of ecosystem response to
perturbation. This framework is shown to provide a new perspective on the
population dynamics of real ecosystems. The formalism is applied to the
problem of the population oscillations of the southern pine beetle.
\end{abstract}

%

\textbf{keywords:} ecosystem perturbation, ecosystem thermodynamics

\section{Introduction}

Most ecosystems are under considerable stress, having been perturbed by
human intrusions including, population reduction or complete species
annihilation, the introduction of foreign species, habitat destruction or
fragmentation, contamination, and general global warming. We are now
witnessing one of the highest rates of species disappearance and ecosystem
collapse over the entire history of life on Earth.

Fortunately, ecosystems can often recover from perturbations and can even
evolve and adapt to new boundary conditions (Norberg et al., 2001). However,
successful recovery depends on the inherent stability of the system, which
is a complex function of the individual interactions among all the
participating species and among species and their environment. Given that
typical ecosystems contain over 3000 species (Polis, 1991), understanding
the nature of this stability, and thus predicting ecosystem response to
perturbation, is far from trivial, but indispensable for obtaining a
quantitative understanding of ecosystem dynamics.

Predicting ecosystem response to perturbation is, therefore, one of the most
scientifically taxing yet important questions of our time. Present ecosystem
dynamics theory is based on empirically inspired but essentially ad hoc
equations incorporating one-body parameters, such as inherent birth and
death rates, and two-body effects of one species population on another
through coefficients representing competition, predator-prey, symbiosis, or
neutral interaction, as well as species-environment effects, incorporated
through parameters such as the ``carrying capacity''. This two-body
``community matrix'' approach, although widely recognized for its usefulness
in revealing the general spectrum of the dynamics of model ecosystems (May,
1974), has had little application to predicting real ecosystem response to
perturbation. This is primarily due to the fact that the dynamical equations
are ad hoc\emph{\ }and that the community matrix is obtained from fits to
time-series population data, and therefore can be expected to be
representative of nature only within the limited range of the available
population data.

A further problem debilitating the community matrix approach is that it is a
two-body approach while species interactions are really of a many-body
nature. These many-body effects are usually absorbed within so called
``environmental factors'' which are included in the dynamical equations as
fitted constants. However, these ``constants'' are not really constant for
perturbed ecosystems and anyhow fail to endow the resultant 2-body equations
with the true dynamics of inherently many-body natural ecosystems.

A clear indication of the failure of traditional theory is the fact that
today's ecosystem health is usually surmised by making painstaking field
counts of the populations of particular species and then, rather
arbitrarily, deciding whether or not to include those species on an ``in
danger of extinction'' list. Such one dimensional and last minute vigilance
of ecosystems is not satisfactory for a number of reasons: first, it fails
to treat the ecosystem as an integrated whole and could thereby miscalculate
the gravity of the situation about to unfold; second, since many ecosystems
have a natural cyclical, or even chaotic, but stable dynamics, it may be
difficult to distinguish normal, but stable, periodic or chaotic behavior
from a dangerous fall toward extinction; third, our human perspective tends
to focus on species in which the individuals are physically large, easily
observable, or likable, but not necessarily those key species that are most
important to the stability of an ecosystem. Most important, however, is the
fact that present ecosystem theory provides no information for designing an
integral solution for arresting an impending catastrophe, other than,
perhaps, suggesting that the endangered specie be protected by law.

There is clearly a need for a more quantitative approach to population
modeling based on fundamental science and measurements that can predict
ecosystem dynamics for regions in population space for which no data exists.
There has been a growing realization that such a quantitative theory of
ecosystems will have a thermodynamic basis (Odum, 1969, Gallucci, 1973,
Swenson, 1989, Michaelian, 2005). The reasons are compelling: First,
thermodynamic laws derive from symmetry principles inherent in nature and
thus are universal, applicable in suitable form to all macroscopic systems
and processes, irrespective of the types of interactions involved. Second,
thermodynamics deals with a much reduced set of macroscopic variables which
can be related with measurable ecosystem variables (e.g. populations)
involved in the dynamical patterns observed in Nature. Third, a number of
stubborn problems and paradoxes existing in traditional ecosystem theory
appear to have a simple resolution in terms of thermodynamic directives
(Swenson, 1989, Michaelian, 2005). The objective of this paper is to
demonstrate that, for ecosystems under constant boundary conditions, a
non-equilibrium thermodynamic framework for the population dynamics can lead
to explicit predictions concerning ecosystem response to perturbation.

In the following section we briefly outline the thermodynamic framework for
treating ecosystems which has been presented elsewhere (Michaelian, 2005).
In section 3 we present a simple model ecosystem and demonstrate how its
population dynamics and stability characteristics are determined by
thermodynamic constraints and criteria relating to energy, work, and mass
flow among the species populations, and with the external environment. In
section 4 we perturb this ecosystem and analyze the response as predicated
on the basis of non-equilibrium thermodynamic formalism. Finally, in section
5, we discuss how our thermodynamic framework may have relevance in
explaining the particular population dynamics observed in a real ecosystem;
the outbreak of sustained population oscillations of the southern pine
beetle.

\section{Thermodynamic Framework}

To avoid misinterpretation, it is prudent to first make a clear distinction
between two existing but fundamentally different thermodynamic frameworks. 
\emph{Equilibrium thermodynamics }deals with isolated systems and the
fundamental state variable governing the evolution of the isolated system
toward the stable \emph{equilibrium} state is the total entropy, $S$. \emph{%
Irreversible thermodynamics} deals with open systems or processes, such as
ecosystems, which exchange energy, work, and mass between component parts
and with the environment. Here, the variable governing the evolution toward
the stable \emph{stationary} state (for constant boundary conditions) is the
time variation of the total entropy of the system, $dS/dt$. Our framework is
based on the latter \emph{irreversible thermodynamics } and we employ only
that part of this framework, known as \emph{classical}, developed by Lars
Onsager (Onsager, 1931) and Illya Prigogine (Prigogine, 1967) that has been
verified empirically.

As for any open system, the time variation of the total entropy of the
ecosystem may be divided into a part due to the internal entropy production
arising from irreversible processes occurring within the ecosystem itself,
and a second part due to the flow of entropy into, or out of, the ecosystem
from the external environment (Prigogine, 1967), 
\begin{equation}
\frac{dS}{dt}=\frac{d_{i}S}{dt}+\frac{d_{e}S}{dt}.  \label{eq:dsdt}
\end{equation}

All macroscopic systems and processes, including ecosystems, are subject to
definite thermodynamic law. The primary among these is the second law of
thermodynamics which states that the internal production of entropy due to
irreversible processes occurring within the system must be positive
definite, 
\begin{equation}
\frac{d_{i}S}{dt}>0.  \label{eq:2ndlaw}
\end{equation}

For the case of ecosystems under the condition of constant external
constraints (see (Michaelian, 2005) for justification of this condition for
a large class of ecosystems) classical irreversible thermodynamic theory
states (Prigogine, 1967) that the system will eventually arrive at a
thermodynamic \emph{stationary state} in which all macroscopic variables,
including the total entropy, are stationary in time, 
\begin{equation}
\frac{dS}{dt}=0.
\end{equation}
Therefore, from (\ref{eq:dsdt}), at the stationary state, 
\begin{equation}
\frac{d_{i}S}{dt}=-\frac{d_{e}S}{dt},  \label{eq:desdis}
\end{equation}
implying from Eq. (\ref{eq:2ndlaw}) that 
\begin{equation}
\frac{d_{e}S}{dt}<0.  \label{eq:desdtlt0}
\end{equation}
Maintaining an ecosystem in a stable thermodynamic stationary state thus
requires a continuous negative flow of entropy into the system. This has
already been emphasized by Schr\"{o}dinger (Schr\"{o}dinger, 1944), but was
first recognized by Boltzmann (Boltzmann, 1886).

The internal entropy production $d_{i}S/dt$ can be written as a sum of
generalized thermodynamic forces $X$ multiplied by generalized thermodynamic
flows $J$ (Prigogine, 1967) (for example, $X_{Q}=$ $\nabla \left( \frac{1}{T}%
\right) $gradient of the inverse temperature, and $J_{Q}=$ heat flow), 
\begin{equation}
\frac{d_{i}S}{dt}=\sum_{\alpha }X_{\alpha }J_{\alpha }.  \label{eq:forceflow}
\end{equation}
The separation of the entropy production into its components of
thermodynamic forces and flows is somewhat arbitrary and can often be chosen
for convenience in resolving the particular problem at hand. However, there
are a number of conditions that must be met for any particular choice. The
first condition is that the product of the force and flow gives units of
entropy production, and the second is that symmetry aspects must be
respected, for example, a scaler force cannot give rise to a vector flow
(Katchalsky and Curran, 1975). We have shown (Michaelian, 2005) that
ecosystem dynamics can be treated consistently within this irreversible
thermodynamic framework by assigning the generalized thermodynamic forces to
the species populations ($X_{\alpha }\equiv p_{\alpha }$) (where $\alpha $
represents the species type) and the generalized flows to the flows of
entropy ($J_{\alpha }\equiv d_{i}S_{\alpha }/dt$) (due to flows of energy,
work, or mass, to or from species $\alpha $, see below).

Within this framework, it was shown (Michaelian, 2005) that the ecological
steady state, prevalent in nature (Goldwasser \& Roghgarden, 1993; Polis,
1991), has the stability characteristics of the thermodynamic stationary
state. In view of this, we have made the formal assertion that the
ecological steady state is just a particular case of the more general
thermodynamic stationary state (Michaelian, 2005).

A further criterion from classical irreversible thermodynamic theory,
considered by Prigogine as the most general result of irreversible
thermodynamic theory, and valid for constant external constraints, is that
the rate of change of the internal entropy production, due to changes in the
generalized forces $X$ (the populations), is negative semi-definite; the 
\emph{general evolutionary criterium} (Prigogine 1967), 
\begin{equation}
\frac{d_{X}\mathcal{P}}{dt}\le 0\verb|   where |\mathcal{P}\equiv \frac{%
d_{i}S}{dt}.  \label{eq:dxpdt}
\end{equation}
Equation (\ref{eq:dxpdt}) implies that, under constant boundary conditions,
all natural changes in the species populations must be in such a manner so
as to reduce the internal production of entropy. This is a powerful
auxiliary criterion on ecosystem response to perturbation and it will be
shown below that this, together with the second law of thermodynamics, and
the fact that a system with constant external constraints must arrive at a
thermodynamic stationary state, effectively determines the population
dynamics that the ecosystem can assume. In this way, we can predict the
actual dynamic response of the ecosystem to perturbation, be it either
toward recovery, toward a new dynamics, or toward extinction.

\section{Model Ecosystem}

We now present our thermodynamic framework for an over simplified but
illustrative 3-species model ecosystem, including up to 3-body interaction
terms. Two of the populations, $p_{1}$ and $p_{2}$, are considered variable,
while the third, $p_{3}$, is fixed, and represents the constant boundary
conditions over the ecosystem; such as the constant supply of nutrients due
to a primary producer species. For example, in our application to the pine
beetle ecosystem, to be detailed in the section 5, $p_{3}$ would represent
the approximately constant population of the pine trees vulnerable to
infection; $p_{1}$ will be taken as the population of the southern pine
beetle; and $p_{2}$ that of its most important natural predator, the clerid
beetle \textit{Thanasimus dubius}.

The total entropy brought into the ecosystem or carried out of it through
one-body transport processes can be written as (Michaelian, 2005), 
\begin{equation}
\frac{d_{e}S}{dt}=\sum_{\alpha =1}^{n}p_{\alpha }\Gamma _{\alpha }^{e},
\label{eq:desdt}
\end{equation}
where the sum is over all $n=3$ species and $p_{\alpha }$ is the population
of species $\alpha $. $\Gamma _{\alpha }^{e}$ represents the average rate of
exchange, or flow, of entropy with the external environment per individual
per unit time of species $\alpha ,$ as a result of energy (including heat),
work, or matter flow (see below).

Similarly, the internal entropy production, including individual production
and exchange of entropy between individuals of the species, may be written
as 
\begin{equation}
\mathcal{P}\equiv \frac{d_{i}S}{dt}=\sum_{\alpha =1}^{n}p_{\alpha }\left[
\Gamma _{\alpha }+\sum_{{\alpha ^{\prime }}=1}^{n}p_{{\alpha ^{\prime }}%
}\Gamma _{\alpha {\alpha ^{\prime }}}+\sum_{{\alpha ^{\prime }},{\alpha
^{\prime \prime }}=1}^{n}p_{{\alpha ^{\prime }}}p_{{\alpha ^{\prime \prime }}%
}\Gamma _{\alpha {\alpha ^{\prime }}{\alpha ^{\prime \prime }}}+O(4)\right]
>0.  \label{eq:disdt}
\end{equation}
The $\Gamma _{\alpha }$ represent the entropy production of species $\alpha $
due to one-body irreversible processes occurring within the individual such
as; photosynthesis, transpiration, respiration, metabolism, etc. The $\Gamma
_{\alpha {\alpha ^{\prime }}}$ represent the entropy production and exchange
due to 2-body interactions between individuals of species $\alpha $ and $%
\alpha ^{^{\prime }}$ (e.g.. those involved in competition, predator-prey,
symbiosis, etc.); $\Gamma _{\alpha {\alpha ^{\prime }}{\alpha ^{\prime
\prime }}}$ correspond to similar but 3-body interactions, and $O(4)$
represents the entropy production due to 4-body and higher order
interactions (for example, those required for the functioning of societies).
The 4-body and higher order n-body terms will be neglected in what follows
since they would normally be small as they require increasingly improbable
(except for social species) n-body localization in space and time.

Equation (\ref{eq:dxpdt}), the general evolutionary criterium, for the time
change in the entropy production due to a change in the generalized forces $%
X $ (the populations) then becomes 
\begin{equation}
d_{X}\mathcal{P}=\sum_{\alpha }dp_{\alpha }\left[ \Gamma _{\alpha }+\sum_{{%
\alpha ^{\prime }}}p_{{\alpha ^{\prime }}}\Gamma _{\alpha {\alpha ^{\prime }}%
}+\sum_{{\alpha ^{\prime }}{\alpha ^{\prime \prime }}}p_{{\alpha ^{\prime }}%
}p_{{\alpha ^{\prime \prime }}}\Gamma _{\alpha {\alpha ^{\prime }}{\alpha
^{\prime \prime }}}\right] \le 0.  \label{eq:dxp}
\end{equation}

The dynamics of the ecosystem can now be determined from equations (\ref
{eq:desdis}), (\ref{eq:desdt}), (\ref{eq:disdt}) and (\ref{eq:dxp}) once the 
$\Gamma $ are specified.

The $\Gamma $'s represent entropy production and flow between individuals of
the species and between individuals and their environment. A general
expression for this entropy flow comes from the Gibb\'{'}s equation and
results from the flow of energy, work, and matter (Prigogine 1967). For
example, the energy per individual per unit time taken in through
photosynthesis $de_{\alpha }$, or the heat $dq_{\alpha }$ per individual per
unit time transported to the external environment, the work done on the
environment per unit time $PdV_{\alpha }$ at constant pressure $P$, and the
matter components (e.g.. nutrients) of type $\beta $ taken in or given out
by species $\alpha $, $dn_{\alpha \beta }$, of chemical potential $\mu
_{\alpha \beta }$, give for the rate of entropy exchange per individual with
the environment, 
\begin{equation}
\Gamma _{\alpha }^{e}=\frac{1}{T}\frac{de_{\alpha }+dq_{\alpha }}{dt}+\frac{P%
}{T}\frac{dV_{\alpha }}{dt}-\frac{1}{T}\sum_{\beta }\mu _{\alpha \beta }%
\frac{dn_{\alpha \beta }}{dt},  \label{eq:gmgm}
\end{equation}
where the temperature $T$ (of the participating individuals) may be
approximated as being constant for the ecosystem (Gallucci, 1973). A similar
expression can be written for the $\Gamma _{\alpha \alpha ^{\prime }}$,
representing the entropy production and exchange between individuals of
species $\alpha $ and $\alpha ^{\prime }$, i.e. in terms of the energy,
work, and matter exchanged due to the 2-body interactions between
individuals of the species. The affect on the entropy flows due to a
simultaneous interaction of a third individual (three-body effects) of
species $\alpha ^{\prime \prime }$ is considered in the parameter $\Gamma
_{\alpha {\alpha ^{\prime }}{\alpha ^{\prime \prime }}}.$

Determining the $\Gamma $'s for a real ecosystem therefore requires the
determination of the flows of energy $de$, heat $dq$, volume $dV$, and mass $%
dn_{\beta }$ of type $\beta $ between individuals of the participating
species and between individuals and the external environment. Such a
determination is possible in principle but obviously difficult in practice.
Some of the experimental details for obtaining these types of flow
measurements between the individuals, and between the individuals and their
environment, can be found in (Gallucci, 1973) and references therein. We
have outlined the empirical determination of the entropy production
coefficient $\Gamma _{\alpha }$ for a plant (Hern\'{a}ndez, 2008,
Hern\'{a}ndez and Michaelian, 2008). The entropy exchange between individual
animals $\Gamma _{\alpha {\alpha ^{\prime }}}$ is more difficult to measure.
However, since much of the entropy flow between individuals is directly
related to their diet, it is plausible, for example, that values for the $%
\Gamma _{\alpha {\alpha ^{\prime }}}$ may one day be obtained through a
chemical analysis of the DNA of the excrement of individuals. Similar
analyses have in fact been performed for real ecosystems, however, with the
focus on energy flow (see for example, Homer et al., 1976).

In the absence of real ecosystem data concerning the production of entropy
and the exchange of entropy, here we generate these coefficients $\Gamma $
for a model ecosystem subject to thermodynamic law with the aid of a genetic
algorithm (Michaelian, 1998). The algorithm begins by randomly generating
sets of initial values for the $\Gamma $'s within fixed ranges. The
algorithm then evaluates the fitness of each set by checking to see if the
population dynamics, as determined by criteria (\ref{eq:disdt}) and (\ref
{eq:dxp}), leads to a viable stationary state in the long time limit, i.e.
one with $\lim_{t\rightarrow \infty }d_{i}S/dt=-d_{e}S/dt$ (as required by
classical irreversible thermodynamics for constant external constraints, see
Eq. (\ref{eq:desdis})) and with $d_{i}S/dt$ large and positive (consistent
with what is known about the natural evolution of biotic systems to ever
higher entropy production regimes (see Prigogine, 1967)). The best sets of $%
\Gamma $'s are selected and evolved through mutation and crossover,
optimizing (maximizing), the fitness function, 
\begin{equation}
\frac{d_{i}S/dt}{d_{i}S/dt+d_{e}S/dt},  \label{eq:fitness}
\end{equation}
which, as required, is large for $d_{i}S/dt$ large and for $d_{i}S/dt\cong
-d_{e}S/dt.$

An example of a set of $\Gamma $'s so obtained is given in Appendix A. Using
this set of $\Gamma $'s and a starting population set of populations ($%
p_{1}=1000,p_{2}=2000$) with $p_{3}$ fixed at $2000$ (the constant external
constraint), and generating infinitesimal variations of the populations $%
dp_{1}$ and $dp_{2}$ at random ($dp_{3}=0$) while only accepting those sets $%
\{dp\}$ which satisfy the thermodynamic criteria of Eqs. (\ref{eq:disdt})
and (\ref{eq:dxp}), leads to the stable cyclic attractor (oscillating)
population dynamics as shown in figure \ref{fig:ca}.

\FRAME{ftbpFU}{3.8354in}{2.6913in}{0pt}{\Qcb{(a) Populations $p_{1}$ (solid
line) and $p_{2}$ (dashed line) as a function of time $k$ for the ecosystem
given in Appendix A in response to various (see text) perturbations. (b) $%
d_{i}S/dt$ (solid line) and $-d_{e}S/dt$ (dashed line) as a function of
time. (c) Trajectory in population space showing the cyclic attractor
dynamics and the effects of 4 distinct perturbations (see text).}}{\Qlb{%
fig:ca}}{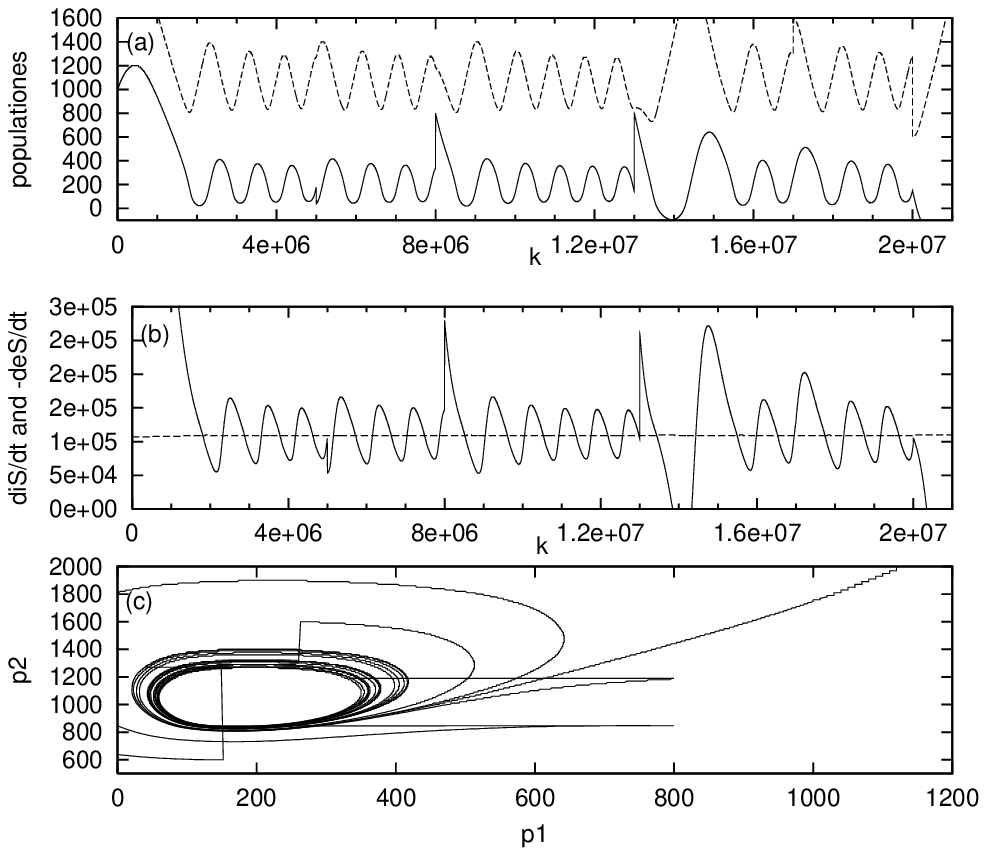}{\special{language "Scientific Word";type
"GRAPHIC";maintain-aspect-ratio TRUE;display "USEDEF";valid_file "F";width
3.8354in;height 2.6913in;depth 0pt;original-width 361.1875pt;original-height
252.75pt;cropleft "0";croptop "1";cropright "1";cropbottom "0";filename
'D:/work/evol/perturb/ca.ps';file-properties "XNPEU";}}

\section{Response to Perturbation}

\label{response}

The response to perturbation obtained under the dictates of the
thermodynamic criteria, Eqs.. (\ref{eq:disdt}) and (\ref{eq:dxp}), once the
ecosystem has arrived at a stationary state, Eq. (\ref{eq:desdis}), are
shown in figure \ref{fig:ca}. This stationary state, for the particular set
of $\Gamma $'s obtained, is a cyclic attractor. The first perturbation,
affected at time $k=5e6$ by reducing the population of $p_{1}$ to $40,$ can
be seen to have little effect on the ecosystem, a full recovery of the
populations is obtained rapidly. The second perturbation affected by
increasing the population of $p_{1}$ to $800$ at time $k=8e6$ also produces
only a small transient effect. However, if the ecosystem is perturbed in the
same manner but at time $k=1.3e7$, population $p_{1}$ goes negative (as does
the internal production of entropy $d_{i}S/dt$) implying the extinction of
the species and the non-viability, thermodynamically, of the ecosystem. The
timing of a perturbation affected on an ecosystem with cyclic attractor
population dynamics therefore appears to be crucial in deciding the fate of
the ecosystem.

Similar results are obtained if population $p_{2}$ is perturbed. A
perturbation of the ecosystem at time $k=1.7e7$ by increasing population $%
p_{2}$ to $1600$ has no long term consequences. However, when population $%
p_{2}$ is reduced to $600$ at time $k=2.0e7$, the internal production of
entropy and the population $p_{1}$ both go negative, the ecosystem again
becomes untenable.

Figure \ref{fig:p1p2} plots the dynamics of the ecosystem in population
space $p_{1}:p_{2}$ for 50 different initial populations. It is apparent
from this figure that the recovery or not of an ecosystem from a particular
perturbation depends on the region in population space into which the
ecosystem is perturbed. Perturbation into the ``regions of danger'' marked
on the figure leads to either one of the populations going negative, or to
the internal production of entropy going negative. Both of these results are
non physical and would foretell the collapse of the ecosystem in nature.
Interestingly, these regions of danger do not necessarily correspond to
regions of small population.

\FRAME{ftbpFU}{3.8458in}{2.6913in}{0pt}{\Qcb{The dynamics in population
space for 50 different initial populations showing that there are ``regions
of danger'' in population space for which a perturbation into these regions
would cause the system to become untenable, either because one of the
populations extinguishes (goes negative) or because the internal production
of entropy becomes negative (non-physical).}}{\Qlb{fig:p1p2}}{p1p2h.eps}{%
\special{language "Scientific Word";type "GRAPHIC";maintain-aspect-ratio
TRUE;display "USEDEF";valid_file "F";width 3.8458in;height 2.6913in;depth
0pt;original-width 520.9375pt;original-height 363.625pt;cropleft "0";croptop
"1";cropright "1";cropbottom "0";filename
'D:/work/evol/perturb/p1p2h.eps';file-properties "XNPEU";}}

We next examine the response of our model ecosystem to permanent changes in
the boundary conditions. Figure \ref{fig:pa1} shows the ecosystem dynamics
under new boundary conditions of $p_{3}$ reduced to $200$, down from its
originally fixed value of $p_{3}=2000$, implying a less negative flow of
entropy into the ecosystem. Without allowing time for the interaction
coefficients $\Gamma $ to evolve in response to the new boundary conditions,
the cyclic attractor then becomes a point attractor as shown in figure \ref
{fig:pa1}. However, as can be seen from this figure, the point attractor is
not a thermodynamic stationary state since the internal production of
entropy is no longer equal to the negative of the external flow of entropy
(equation (\ref{eq:desdis}) is no longer satisfied). The ecosystem fitness
function, Eq.. (\ref{eq:fitness}), is no longer at a local maximum value and
the system, given time, would evolve its entropy production and exchange
coefficients (the set $\Gamma $) until reaching a new stationary state where
the production and external flow of entropy are once again equal. Note that
in the perturbed state with the new external constraint, $p_{3}=200,$ the
same small perturbation of reducing the population $p_{1}$ to 40 at time $%
k=5e6$, which did not have any lasting affect on the ecosystem previously
(Fig. \ref{fig:ca}), now results in the collapse of the ecosystem since it
moves it into the non-physical, thermodynamically prohibited, regime of
negative internal entropy production (figure \ref{fig:pa1}).

\FRAME{ftbpFU}{3.8354in}{2.6913in}{0pt}{\Qcb{Ecosystem dynamics for the case
where the fixed boundary conditions have been changed from $p_{3}=2000$ to $%
p_{3}=200$. The population dynamics becomes that of a point attractor.
However, the system is not in a thermodynamic stationary state since $%
d_{i}S/dt\neq -d_{e}S/dt$. A subsequent perturbation of $p_{1}$ to $40$ at $%
k=5e6$, which did not affect the stability previously (see figure 1), now
leads to $d_{i}S/dt$ going negative, violating the 2nd law. The
vulnerability of the ecosystem has thus been increased by reducing the
external constraint of $p_{3}$.}}{\Qlb{fig:pa1}}{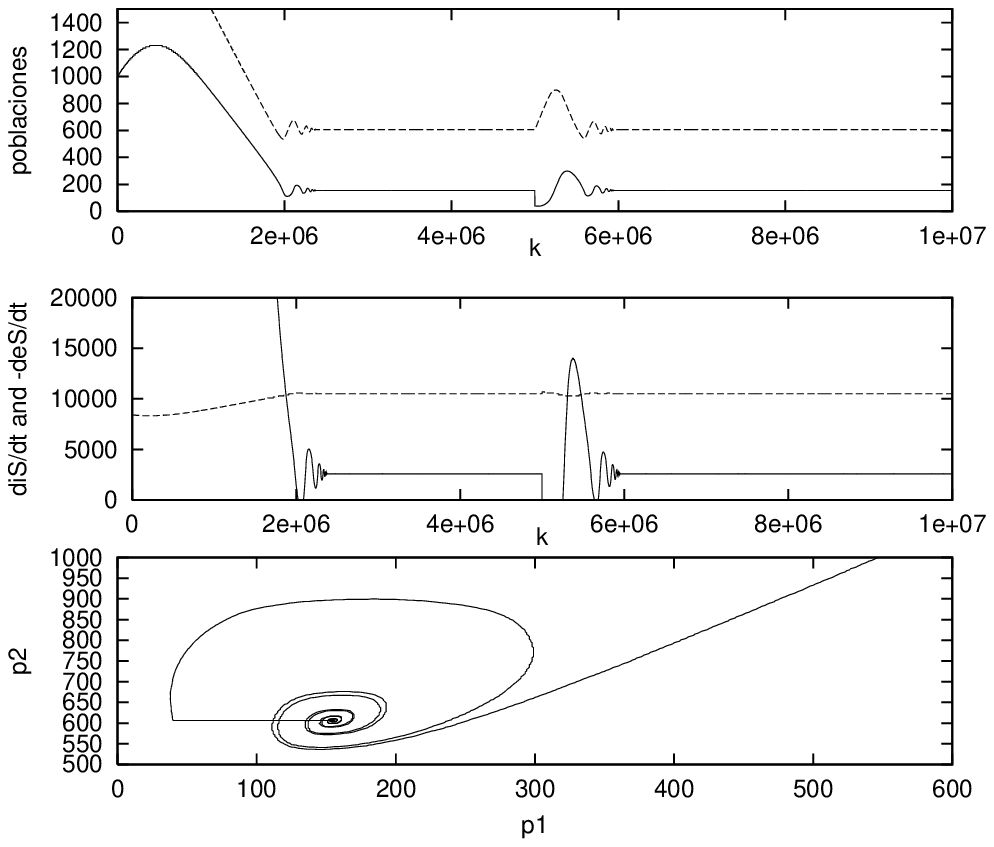}{\special{language
"Scientific Word";type "GRAPHIC";maintain-aspect-ratio TRUE;display
"USEDEF";valid_file "F";width 3.8354in;height 2.6913in;depth
0pt;original-width 361.1875pt;original-height 252.75pt;cropleft "0";croptop
"1";cropright "1";cropbottom "0";filename
'D:/work/evol/perturb/pa1.ps';file-properties "XNPEU";}}

Figure \ref{fig:ca4} shows the opposite effect of increasing the flow of
negative entropy into the ecosystem, obtained by increasing the value of the
fixed external condition to $p_{3}=2200$. Without allowing time for the
interaction coefficients $\Gamma $ to evolve, the dynamics remains that of a
cyclic attractor but the orbit of the attractor increases significantly,
bringing the population $p_{1}$ very close to zero at one point in its
orbit. A slight perturbation of population $p_{2}$ at time $k=5e6$ is
sufficient to cause the population $p_{1}$ to pass through zero and thereby
cause the collapse of the ecosystem. We believe that this is a possible
thermodynamic explanation of the ``enrichment paradox''(Rosenzweig, 1971);
contrary to naive expectation, an increase in the inflow of nutrients is
often observed to make an ecosystem more vulnerable to perturbation. This
will be considered in detail in a forthcoming paper (Alonso, 2007).

\FRAME{ftbpFU}{3.8354in}{2.6913in}{0pt}{\Qcb{Ecosystem dynamics for the case
where the fixed boundary condition has been increased from $p_{3}=2000$ to $%
p_{3}=2200$. In this case, the population dynamics is that of a cyclic
attractor with a larger orbit, bringing $p_{1}$very close to zero at times.
A subsequent small perturbation of $p_{2}$ to $800$ at $k=5e6$ leads to $%
d_{i}S/dt$ going negative and population $p_{1}$passing through zero. We
believe that this is the thermodynamic origin of the ``enrichment paradox''
(Rosenzweig 1971).}}{\Qlb{fig:ca4}}{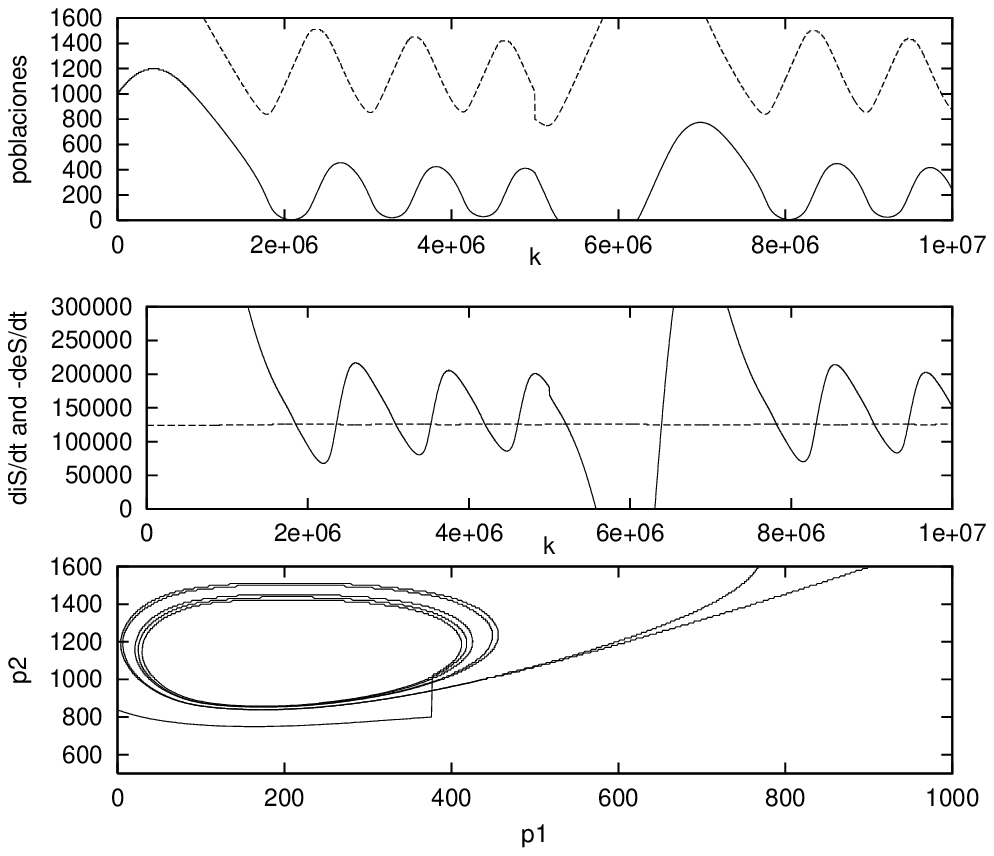}{\special{language "Scientific
Word";type "GRAPHIC";maintain-aspect-ratio TRUE;display "USEDEF";valid_file
"F";width 3.8354in;height 2.6913in;depth 0pt;original-width
361.1875pt;original-height 252.75pt;cropleft "0";croptop "1";cropright
"1";cropbottom "0";filename 'D:/work/evol/perturb/ca4.ps';file-properties
"XNPEU";}}

We have also verified (Alonso, 2007) that had the set of $\Gamma $'s been
chosen such that the thermodynamic stationary state corresponded to a point
attractor in population space (sometimes referred to as an \textit{%
equilibrium state }in the ecological literature), then increasing the value
of the fixed external condition $p_{3}$ leads to population oscillations of $%
p_{1}$ and $p_{2}$. This may have relevance to the sudden outbreak of large
population oscillations of certain insect pathogens, such as the southern
pine beetle, as described in the following section.

\section{Southern Pine Beetle Population Oscillations}

The Southern Pine Beetle, indigenous to the southern and south eastern
states of the United States and to the northern parts of Mexico, is one of
the pathogens of most concern to humans as it can severely affect the
economic value of a forest stand. Its population dynamics often exhibits
infestation outbreaks of an oscillatory nature. This dynamics has been
extensively studied with a view to understanding its origin and thereby to
controlling the pest (Turchin et al., 1999). In the wild state of natural
forests, the pine beetle populations are small and relatively constant and
the insect may be considered as an integral component of a robust ecosystem,
even helping to promote normal ecosystem succession (Natural Resources
Canada, 2003). However, the last century has seen an almost complete
replacement of the natural mixed forest stands with homogeneous stands of
particular species of greater economic value to humans, such as soft wood
pine. This has led to forests more vulnerable to pine beetle attack with the
result that millions of hectors of pine trees can be devastated by a single
beetle outbreak (Aukema et al., 2008). The outbreak is often preceded by a
debilitating event affecting the forest such as a drought, fire, or flood
(McNulty et al., 1998).

The details surrounding an outbreak of a sustained pine beetle attack are
well known. Normally the pine beetle population is relatively small and
constant with the beetle preying only on the vulnerable weaker trees of a
stand. The vulnerable pine trees provide nourishment and protection for the
pine beetle which buries under the inner bark of the tree to deposit its
eggs in bored out gallerias. The beetles carry a fungus which quickly
establishes itself within the tree and spreads out to cut off the water and
nutrient irrigation of the tree. The vulnerable tree normally dies within a
year of first attack. Healthy trees efficiently repel the pine beetle attack
by emitting enough resin in the beetle galleries to dissuade further
colonization (Blinka, 2007) and induce beetle mortality through a number of
effects including crystallization of the sap around beetle larva.

Unlike vulnerable pine trees, healthy pine trees, therefore, cannot normally
be considered as contributing to the primary host species of the pine
beetle. If, however, there occurs a debilitating event, such that the
density of vulnerable pine trees in a given area becomes great enough, then
the local population of pine beetles may swell to such a point that there is
now a large enough number of beetles on each tree ($\sim 200/$m$^{2})$ that
even the healthy trees cannot cope and become vulnerable (Blinka, 2007).

It has long been debated as to whether exogenous (originating externally) or
endogenous (originating from within) factors contribute most to the observed
population oscillations of the pine beetle. Turchin (Turchin, 2003) has
argued for a synthetic approach, suggesting that both enogenous and
endogenous factors can be important. For example, while it is known that
particularly cold winters can reduce and even end an outbreak, otherwise
very little correlation of climate on population dynamics has been found
(Turchin et al., 1991). This fact has lead to a search for possible
endogenous effects to explain oscillations in the population density of the
pine beetle.

One endogenous effect studied by Turchin et al. is a time delay in
density-dependent affect of a predator on the pine beetle. The effect of
introducing a time delay into the normally stabilizing density dependent
terms of Lotka-Volterra type equations was first studied theoretically by
May (1974) for the systems of vegetation-herbivore, and
vegetation-herbivore-predator. For the vegetation-herbivore system, May
found that a time delay in the normally stabilizing density term of the
logistic equation could lead to limit cycles (stable population
oscillations) if the characteristic growth rate of the herbivore was small
compared to the time delay in the regulatory mechanism (stabilizing term of
the logistic equation related to the density of the herbivore). When a
predator was included to prey on the herbivore, May found it much more
difficult to obtain oscillatory solutions to the model equations.
Nevertheless, through carefully controlled field experiments, Turchin et al.
(1999) verified that there indeed exists a delayed density-dependent affect
of predators on the pine beetle which could thus contribute to the
population oscillations.

The essence of the pine beetle ecosystem can be reduced to three strongly
interacting species; the primary production species (the vulnerable pine
tree), the herbivore (the pine beetle itself) and a predator species (the
clerid beetle, the most important natural predator of the pine beetle). This
reduction of the ecosystem to only three species is convenient (but not
necessary) to demonstrate how our formalism, based on the application of
general thermodynamic criteria concerning entropy production and exchange,
can describe the dynamics of the pine beetle ecosystem.

In section \ref{response} we have shown that if the interactions (exchange
and production of entropy) between individuals of the different species of
an ecosystem is of n-body nature (including higher than 2-body terms) then a
stable point attractor or stable limit cycle dynamics can be obtained after
the system relaxes to its thermodynamic stationary state (see figure \ref
{fig:ca1}). Consider now such a stable ecosystem in its stationary state,
being either a point attractor or stable limit cycle in population space.
The initial constant boundary conditions for this ecosystem correspond to
that segment $p_{3}$ of the total population of pine trees within a given
area that are vulnerable to attack while the ecosystem is enjoying normal,
stable conditions. If the population $p_{3}$ of the vulnerable trees then
increases suddenly due to a debilitating event affecting the forest, then
the beetle population $p_{1}$ and the Clerid beetle predator population $%
p_{2}$ start to oscillate with potentially much greater amplitude (Fig. \ref
{fig:ca4}).

The interaction (production and exchange of entropy) of the pine beetles
with the pine trees and its predator is non-linear in their respective
populations and the physical restrictions on the total entropy production
(the 2nd law) and on the sign of the rate of change of the production of
entropy (Prigogine's general evolution criterium) together force this
oscillatory population dynamics. However, this is a perturbed ecosystem, one
which is no longer in a thermodynamic stationary state, as could be verified
in the field by demonstrating that the production of positive entropy has
temporarily increased and is no longer equal to that of the inflow of
negative entropy (see Eq. \ref{eq:desdis} and figure \ref{fig:ca4}).

This explanation of the population oscillations is in accord with the
following known facts regarding pine beetle population dynamics; 1) the pine
beetle populations are not always oscillating but can be relatively constant
in healthy forest stands, 2) oscillations usually begin after a particular
debilitating event which causes a rapid increase in the number of vulnerable
trees, the primary resource available to the beetle, 3) the amplitude of the
population oscillations are correlated with the \emph{increase} in the
amount of vulnerable trees in a given area, or, in other words, to the
density of nominally healthy pine trees in the area, 4) it is known that
perturbed ecosystems (including diseased systems) have temporarily increased
total entropy production (Schneider and Sagan, 2005), 5) it is contingent on
both exogenous (debilitating events) and endogenous (thermodynamic laws)
factors.

The results presented above suggest a possible means for controlling a pine
beetle outbreak. Figure \ref{fig:p1p2} shows that there are regions in
population space for which perturbation into these regions leads to eventual
collapse of the ecosystem. Once the entropy flow coefficients $\Gamma $ have
been determined, the thermodynamic formalism presented here determines these
regions of danger. It is then only a matter of perturbing one of the
populations into these regions of danger to obtain the desired extinction of
the outbreak. This could be obtained by augmenting, or reducing, the
population of one of the involved species at the indicated point in the
population cycle.

\section{Summary and Conclusions}

Acknowledging that ecosystems, like all macroscopic processes, are subject
to definite thermodynamic law, we have demonstrated that under constant
external constraints, the response of ecosystems to perturbation can, in
principle, be predicted. The thermodynamic criteria which direct the
dynamics come from non-equilibrium thermodynamic theory. They are; 1) the
system must eventually arrive at a thermodynamic stationary state, Eq. (\ref
{eq:desdis}), 2) the internal production of entropy must be positive
definite, in accord with the second law of thermodynamics, Eq. (\ref
{eq:2ndlaw}), and 3) any natural change in the populations must be in such a
manner so as to reduce the internal production of entropy of the entire
system, Prigogine's general evolution criterion, Eq. (\ref{eq:dxpdt}).

In the absence of data on the entropy production and exchange of real
ecosystems, we considered a simple model ecosystem generated by evolving
interaction coefficients (representing entropy production and exchange)
through selection with a fitness function favoring the thermodynamic
criteria identified above. We then studied the response of this model
ecosystem to perturbation of the populations under the same thermodynamic
criteria. We found that there exists regions in population space for which
perturbation into these regions leads to the eventual extinction of one or
more of the species, or to a negative internal production of entropy. The
latter violates the second law of thermodynamics and would lead to ecosystem
collapse since physical maintenance processes require positive production of
entropy. An important finding is that these regions in population space are
a general feature of the thermodynamic framework and do not necessarily
correspond to regions of small population. Assigning species to ``in danger
of extinction'' lists, solely on the basis of the smallness of their
populations may therefore not be an effective conservation policy. Our
proposed approach, based on thermodynamic criteria, predicts the dynamics
over all of population space and thus leads to quantitative elements for
providing more informed policy statements for responding to ecosystem
perturbation.

Increasing or decreasing the negative flow of entropy (natural resources)
into the ecosystem has the effect of increasing or decreasing respectively
both the amplitude of the orbit of the attractor in population space and the
internal production of entropy of the system. In either case, this results
in a more vulnerable ecosystem since the populations pass closer to zero or
the internal production of entropy may more easily become negative
respectively. We believe that this result is a thermodynamic explanation of
the ``enrichment paradox''.

We applied our thermodynamic framework to a reduced ecosystem consisting of
the southern pine beetle, pine trees, and the beetles most important
predator. We have shown that the population oscillation of the southern pine
beetle may be viewed within this thermodynamic framework as resulting from
an increase in the inflow of resources (negative entropy) into a system
which has a non-linear dependence of the production and flow of entropy on
the species populations. Our thermodynamic description is contingent on the
initial conditions known to precede an attack (the increase in the density
of the vulnerable trees), predicts pine beetle population oscillations for
perturbed ecosystems, and predicts constant (or small amplitude
oscillations) populations for unperturbed ecosystems. The model is
consistent with the synthetic view that oscillations of the pine beetle
populations are a result of both exogenous factors (the initial debilitating
event causing an increase in the number of vulnerable pine trees) and
endogenous factors (thermodynamic laws and non-linear coupling of entropy
flows).

We proposed a method of control of the pine beetle by first delimiting these
regions of danger, and then perturbing the populations into these regions.

Ecosystems are composed of many thousands of interacting species and the
details of the dynamics is, undoubtedly, significantly more complicated.
However, our thermodynamic framework can be straightforwardly applied to a
much larger and more complex ecosystem simply by measuring all the entropy
production and exchange coefficients $\Gamma $ for all the species involved.
Work in this direction is underway (Hern\'{a}ndez, 2008).

In conclusion, as for all macroscopic process, ecosystems are subject to
definite thermodynamic law. For constant external constraints, these laws
are sufficient to determine ecosystem response to perturbation. Our analysis
of the population dynamics based on thermodynamic law and the formulation of
the interaction coefficients in terms of physical and measurable quantities
(the production and exchange of entropy) is a step toward a more
quantitative theory of ecosystems.

\section{Acknowledgments}

K.M. is grateful for the financial support of DGAPA-UNAM project number
IN-118206, and for financial assistance while on a 2007 sabbatical leave.
The hospitality afforded during this leave by the Faculty of Science,
Universidad Autonoma del Estado de Morelos, Mexico, is greatly appreciated.

\newpage \noindent {\Large \textbf{Appendix A}}\newline
\setcounter{equation}{0} \renewcommand{\theequation}{B.\arabic{equation}} \ 
\newline
The following set of species interaction (entropy production and exchange)
coefficients (see Eqs. (\ref{eq:desdt}) and (\ref{eq:disdt})), for an $n=3$
species ecosystem and including up to 3-body terms, were used for
calculating the dynamics of figures \ref{fig:ca} to \ref{fig:ca4}; 
\begin{table}[h]
\begin{tabular}{lll}
$\Gamma^e_{1}= 1.1721$ & $\Gamma^e_{2}= 0.4710$ & $\Gamma^e_{3}= -55.000$ \\ 
$\Gamma_{1}= 0.6277$ & $\Gamma_{2}= 8.787$ & $\Gamma_{3}= 1.1884$ \\ 
$\Gamma_{11}= -0.01616$ & $\Gamma_{22}= -0.04886$ & $\Gamma_{33}= 0.020869$
\\ 
$\Gamma_{12}= 0.0372$ & $\Gamma_{21}= 0.0222$ & $\Gamma_{32}= 0.012619$ \\ 
$\Gamma_{13}= 0.01617$ & $\Gamma_{23}= 0.000339$ &  \\ 
$\Gamma_{111}= 0.000448$ & $\Gamma_{222}= -0.0000078$ &  \\ 
$\Gamma_{122}= -0.0000574$ & $\Gamma_{211}= -0.0002856$ &  \\ 
$\Gamma_{112}= -0.0001446$ & $\Gamma_{221}= 0.0002881$ & 
\end{tabular}
\end{table}
\newline
These coefficients were obtained by evolving various initial random sets of
coefficients, selected from within a finite range, using a genetic algorithm
with a fitness function favoring a stationary state, i.e. $%
d_{i}S/dt=-d_{e}S/dt$ and $d_{i}S/dt$ large in the long-time limit, (ie. at
the stationary state). Although our 3 species ecosystem is artificial, the
coefficients have characteristics of what would be expected of a real
ecosystem. For example, the coefficient of entropy flow with the external
environment, $\Gamma _{3}^{e}$, representing the fixed external constraints,
is of large negative value. The one-body internal entropy production
coefficients $\Gamma _{1}\dots \Gamma _{3}$ are all positive (respecting the
2nd law of thermodynamics) and of large value with respect to the 2-, and
3-body coefficients. The second law for the total internal entropy
production, Eq.(\ref{eq:2ndlaw}), is also respected over a large region of
population space (see figure \ref{fig:p1p2}).

\end{document}